\documentclass[10pt,twocolumn]{article}

\usepackage{fourier}
\usepackage[scaled]{helvet}
\usepackage[T1]{fontenc}
\RequirePackage{times}

\usepackage[comma,authoryear,sort]{natbib}

\usepackage[margin=.6in]{geometry}
\usepackage{setspace}
\usepackage{amsmath}
\usepackage{amssymb}
\usepackage{amsthm}
\usepackage{amsfonts}
\usepackage{mathrsfs}
\usepackage{array}
\usepackage{rotating}
\usepackage{graphicx}
\usepackage[colorlinks=true, allcolors=black]{hyperref}
\usepackage{url}
\usepackage{caption}
\usepackage{subfigure}
\usepackage{bbm}

\title{\sf \textbf{\textsc{forensic science and how statistics can help it}}\\
\small\it Evidence, Hypothesis testing, and Graphical models
}
\author{Xiangyu Xu and Giuseppe Vinci\\
{\small  Department of Applied and Computational Mathematics and Statistics, University of Notre Dame}}

\date{}

\begin{document}
\maketitle
\begin{abstract}
The persistent issue of wrongful convictions in the United States emphasizes the need for scrutiny and improvement of the criminal justice system. While statistical methods for the evaluation of forensic evidence, including glass, fingerprints, and DNA, have significantly contributed to solving intricate crimes, there is a notable lack of national-level standards to ensure the appropriate application of statistics in forensic investigations. We discuss the obstacles in the application of statistics in court, and emphasize the importance of making statistical interpretation accessible to non-statisticians, especially those who make decisions about potentially innocent individuals. We investigate the use and misuse of statistical methods in crime investigations, in particular the likelihood ratio approach. We further describe the use of graphical models, where hypotheses and evidence can be represented as nodes connected by arrows signifying association or causality. We emphasize the advantages of special graph structures, such as object-oriented Bayesian networks and chain event graphs, which allow for the concurrent examination of evidence of various nature.
\end{abstract}


\section{Introduction}
\label{sec: Introduction}
Innocent individuals being wrongly convicted of serious crimes is a recurring occurrence in the United States. There has been a significant interest on wrongful convictions \citep{gross2008convicting} and, according to \cite{NRE2022report}\footnote{A project of the University of California Irvine Newkirk Center for Science \& Society, University of Michigan Law School and Michigan State University College of Law.}, 233 cases of exoneration have occurred in the U.S. in 2022, among which 44 cases involved false or misleading forensic evidence.

U.S. trial courts handle a wide variety of cases, which can be broadly categorized into \textit{civil cases} and \textit{criminal cases}. Civil cases typically revolve around conflicts related to finances, property, or constitutional rights. These disputes may be initiated by an individual, corporation, or government agency through a civil lawsuit. Criminal cases pertain to the accusation of breaching federal criminal statutes and only the government can initiate the criminal proceedings. The court systems and the rules for handling civil and criminal cases vary between states, but for criminal cases they are similar to the federal rules \citep{federalrulesofcriminalprocedure}.

\begin{figure*}[t]
    \centering
    \includegraphics[width=1\textwidth]{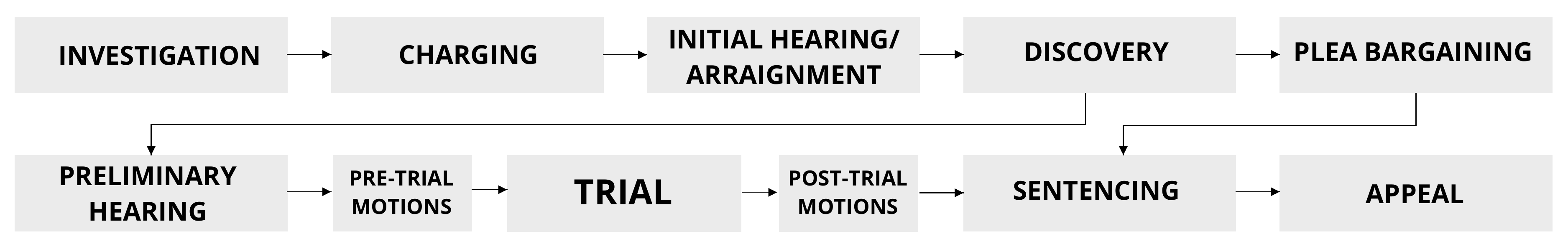}
    \caption{\small\textbf{Federal criminal process.} Forensic evidence and statistics play crucial roles throughout the whole process.}
    \label{fig: Federal criminal process}
\end{figure*}

The general steps of a federal criminal process in the U.S. \citep{offices2023steps} are listed in Figure~\ref{fig: Federal criminal process}.
In the \textit{investigation} step, agencies such as police departments, the Federal Bureau of Investigation (FBI), the Drug Enforcement Administration (DEA), Bureau of Alcohol, Tobacco, Firearms and Explosives (ATF), the U.S. Secret Service (USSS), and the Homeland Security Investigations (DHS/HSI), employ criminal investigators to gather information for the U.S. attorneys, the \textit{prosecutors}, within their designated areas. Upon reviewing the information provided by investigators and conducting interviews with relevant individuals, the prosecutor assesses whether to submit the case to a grand jury of about 16--23 members. For potential felony charges, the grand jury is presented the evidence, witnesses' testimony, along with a summary of the case provided by the prosecutor. The members of the jury confidentially vote to determine if there is sufficient evidence to formally \textit{charge} an individual, the \textit{defendant}, with a crime. Typically within a couple of days, the defendant is brought before a magistrate judge for an \textit{initial case hearing}, where the judge decides whether the defendant will be detained in custody or released until the trial. Then, the prosecutor and the defense attorney engage in discussions with witnesses and analyze the available evidence (\textit{discovery}). The prosecutor and the defendant can both avoid the trial by offering a \textit{plea} deal or pleading guilty, respectively. Otherwise, after an optional \textit{preliminary hearing} where the prosecutor shows that there is enough evidence to prosecute the defendant, the lawsuit then proceeds to the trial stage, which consists of \textit{pre-trial motions}, the \textit{trial}, and \textit{post-trial} motions. The \textit{trial} is a formal procedure in which the prosecutor seeks to convince the jury of the defendant's guilt, whereas the defendant, aided by an attorney, presents their perspective, both through witnesses and evidence. The jury then assesses whether the defendant is innocent or guilty based on the presented evidence. The defendant returns to the court to be \textit{sentenced} by a judge usually a few months after being found guilty. If the defendant believes there was an unjust verdict or the sentence was overly severe, the defendant has the opportunity to \textit{appeal} to the Circuit Court. The defendant is also allowed to appeal the decision by the Circuit Court to the U.S. Supreme Court in Washington, D.C., the highest appellate court in the American court system that can make a final decision about the defendant's appeal.

During the federal criminal process, examiners need to cope with several uncertainties, especially when interpreting forensic evidence. Statistics can help quantify and systematically reduce these uncertainties. In this paper we describe the important role played by statistics in forensic science. In Section~\ref{sec: Evidence}, we describe different types of evidence commonly used in forensic investigations. In Section~\ref{sec: The challenges of promoting statistics in court}, we discuss the obstacles in the application of statistics in court. In Sections~\ref{sec: Likelihood ratio framework} and~\ref{sec: Graphical models}, we present the likelihood ratio approach and graphical models, which are the two major statistical frameworks used in court. Finally, in Section~\ref{sec: Conclusion} we provide some final remarks on the important role played by statistics in forensic science.

\section{Evidence}
\label{sec: Evidence}
Criminals in the ancient world could easily avoid punishment due to limited standardized forensic practices. These practices were typically limited to forced confession, witness statements, and arguments between prosecutors and defenders \citep{schafer2008ancient}, which were subject to errors and coercion. Over the centuries, forensic science has evolved into a broad field that now offers common tools in criminal and civil law.

\begin{figure*}[t]
    \centering
    \includegraphics[width=1\textwidth]{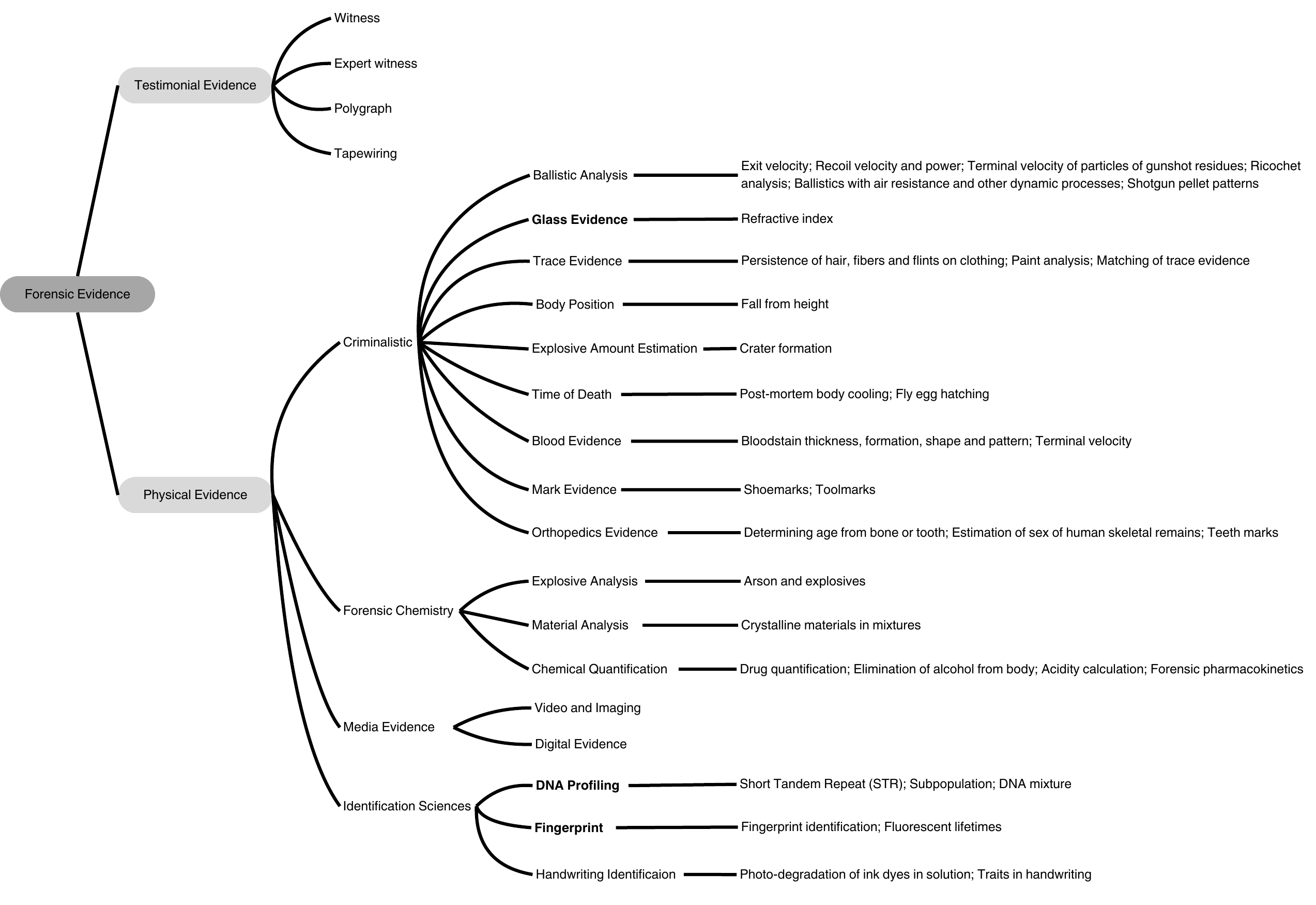}
    \caption{\small\textbf{Mind map of forensic evidence.} Forensic evidence can be categorized into testimonial and physical evidence. Physical evidence is categorized into criminalistics, forensic chemistry, media evidence, and identification science.
    }
    \label{fig: forensic evidence}
\end{figure*}

Forensic evidence can be categorized into testimonial and physical evidence. \textit{Testimonial evidence} is one of the oldest types of evidence in human history. The word ``testimony'' and ``testify'' both come from the Latin word \textit{testis}, which refers to the concept related to a disinterested third-party witness \citep{harper2001online}. Especially in ancient societies, with the lack of standardized forensic practices, the results of trials relied heavily on witness statements \citep{schafer2008ancient}. \textit{Physical evidence} can be categorized into criminalistic, forensic chemistry, media evidence, and identification science, according to the 19th International Forensic Science Symposium (October 2019, France) sponsored by Interpol \citep{houck201919th}. Based on this categorization and several other sources \citep{adam2011essential,fisher2009introduction,saferstein2004criminalistics}, we built a mind map (Figure~\ref{fig: forensic evidence}) that categorizes the various types of forensic evidence commonly used in court.

In the rest of this section, we provide details about common types of physical evidence, in particular fingerprints (Section~\ref{sec: Fingerprint}), glass (Section~\ref{sec: Glass}), and DNA (Section~\ref{sec: DNA}), which will be discussed further in subsequent sections on statistical methods.

\subsection{Fingerprints}
\label{sec: Fingerprint}
Fingerprints were used in court for the first time in 1892 to identify Francisca Rojas as the murderer of her two children in Buenos Aires, Argentina \citep{yard1990fingerprint, beavan2002fingerprints, lambourne1984fingerprint, barnes2010fingerprinthistory}. Fingerprint friction ridge details are typically categorized into three hierarchical levels \citep{jain2006pores}. The first level involves the overall pattern of the ridges, including loops, whorls, and arches. While this level cannot be used for individualization, it can help to exclude suspects from consideration \citep{ashbaugh1999quantitative}. The second level of detail involves minutiae points, also known as \textit{Galton characteristics}, which refer to specific paths of the friction ridges. This level has sufficient discriminating power \citep{pankanti2002individuality,stosz1994automated} and was first studied probabilistically by Francis Galton \citep{galton1892finger}. The third level of detail involves the intrinsic ridge shapes and relative pore locations \citep{ashbaugh1999quantitative}. Although latent fingerprint resolution may not always be high enough to extract level-3 details, these features can be helpful for qualitative examination as they are also permanent, immutable, and unique. According to a study on pore identification conducted by Edmond Locard in 1912, 20 to 40 pores should be sufficient to identify a person \citep{ashbaugh1999quantitative}.

\subsection{Glass}
\label{sec: Glass}
Glass evidence may come from various sources at several crime scenes, for example, broken windows in a burglary, crashed headlight or windshield in a car accident, bottles in an assault, and containers in a domestic violence case or in a laboratory accident \citep{curran2003statistical}. Glass evidence is routinely used in the reconstruction of events, for example, to link an individual to the crime scene \citep{curran2000forensic}. This is accomplished by comparing the average \textit{refractive index} (RI) of the glass fragments recovered from the suspect with one of the fragments sampled from the crime scene. The RI of a glass fragment is measured as follows: first, the fragment is submerged in silicon oil; then, the oil is warmed up gradually until the glass fragment becomes invisible, in which case glass and oil have same RI \citep{ojena1972precise}. The RI of the oil increases linearly with the temperature, and this relationship can be used to measure its RI, and thereby the RI of the glass fragment. A $t$-test can be used to test whether the RIs of the two glass samples are different.

\subsection{DNA}
\label{sec: DNA}
DNA (deoxyribonucleic acid) analysis in forensic investigations has quickly emerged as an indispensable tool in the identification of suspects since its first introduction to forensic science in the mid 1980s \citep{murphy2018forensic,doj2017advancingdna,johnson2004post}. DNA evidence has served as a significant breakthrough in several cases, such as the ``Green River Killings'' in 2001, which had remained unsolved for two decades, despite extensive efforts by law enforcement and a \$15 million investigation \citep{doj2017advancingdna}.

The primary technology for quantitative analysis of DNA evidence is known as \textit{short tandem repeat} (STR). This method focuses on the short sequences of base pairs that occur at specific \textit{loci} along DNA strands \citep{butler2005forensic, adam2011essential}. For example, there are repeated sequences of the base pair ``GATA'' on chromosome 16 in humans at locus \textit{D16S539} \citep{STRFact}. Each distinct sequence is referred to as an \textit{allele}. An example of an allele is ``GATAGATAGATAGATAGATA'', which corresponds to ``allele 5.'' To establish the uniqueness of DNA, information about the alleles at multiple loci, which defines a person's \textit{genotype}, is required. In the U.S., law enforcement utilizes a method based on 20 STR loci, which is at the foundation of the Combined DNA Index System (CODIS) \citep{adam2011essential,FBICODIS,karantzali2019effect}.

Data extracted from a DNA sample can be graphically represented by an electropherogram, where peaks correspond to specific loci and alleles. The presence of a peak at a given location indicates the presence of the corresponding allele type in the sample. Additionally, the height or area of the peaks can be utilized to estimate the proportion of DNA contributed by each individual in a mixed sample (Section~\ref{sec: DNA mixture likelihood framework} and Section~\ref{sec: DNA mixture graphical models}), as there is typically a reasonable correlation between these attributes \citep{buckleton2007towards}.

\subsection{Other types of evidence}
\label{sec: Other types of evidence}
Apart from fingerprints, glass, and DNA evidence, there are a number of other types of evidence used in court, as shown in Figure~\ref{fig: forensic evidence}. These include projectiles, bloodstains, post-mortem body temperature, and body position, which can help with crime scene reconstruction, and orthopedic evidence and mark evidence, which can help with the identification of murder weapon or human identity. Moreover, in the past two decades, due to the swift advancement in computer science technology, digital forensic evidence has transitioned from being a relatively unknown practice to a crucial component of numerous investigations \citep{garfinkel2010digital}. A variety of information can be gathered from suspects' electronic devices, including laptops and mobile phones which contain information such as chat history, browser history, media files, and location history. This digital information can play a significant role in almost any criminal investigation, from minor offenses to cybercrime, organized crime, and even acts of terrorism \citep{aarnes2017digital}.

\section{Obstacles in the application of statistics in court}
\label{sec: The challenges of promoting statistics in court}
The application of statistics in criminal and civil proceedings has a lengthy historical background that has been extensively documented \citep{gastwirth2012statistical,kadane2008statistics,koehler1992probabilities,vosk2014forensic,fenton2016bayes}. The first recorded case in which a statistical analysis was submitted as evidential support is the ``Howland case'' (\textit{Robinson v.~Mandell}) in 1868. Over a century later, there still exist obstacles in the way to the appropriate application of statistics in court. According to \cite{morrison2022advancing}, forensic science still heavily relies on subjective judgment and human perception, which are non-transparent, prone to cognitive bias, and lack of empirical validation. Thus, there is a need for a move towards the rigorous application of statistical methods for the analysis and interpretation of forensic evidence.

In 2009, the National Research Council (NRC) published a report titled \textit{Strengthening Forensic Science in the United States: A Path Forward}, which emphasized the need of national-level standards regarding statistics-related issues including but not limited to interpreting scientific data, dealing with uncertainty and bias, and reporting results \citep{cole2020probabilistic,national2009strengthening}. Besides the U.S., other countries also have similar guidelines for assessing and interpreting forensic evidence, e.g. the United Kingdom \citep{puch2012assessing, jackson2015case}. 

In the rest of this section, we discuss the major challenges of promoting statistics in court: the \textit{misuse of statistics}, especially by prosecution and defense teams (Section~\ref{sec: The misuse of statistics}); \textit{cognitive bias} arising from human perception and subjective judgement (Section~\ref{sec: Cognitive bias}); and other limitations (Section~\ref{sec: Other limitations}), including the \textit{limited understanding} of statistical concepts by laypeople, e.g. members of the jury, and the difficulty in applying \textit{statistical models}.

\subsection{Misuse of statistics}
\label{sec: The misuse of statistics}
On December 13, 1996, lawyer Sally Clark found her 11-week-old son Christopher dead. Two years later, on January 26, 1998, her 8-week-old son Harry was also found dead. The autopsies concluded that the deaths were caused by suffocation, and Sally Clark was convicted to murder on November 9, 1999. However, after serving three years in jail, she was released after the success of a second appeal \citep{byard2004unexpected}. Sally Clark's wrongful conviction in 1999 was due to the \textit{misuse of statistics}, specifically the ``violation of the product rule'' and the so-called ``prosecutor's fallacy.'' During the trial in 1999, the defense argued that the two children had died of sudden infant death syndrome (SIDS), but the expert witness paediatrician Roy Meadow claimed that such explanation was extremely unlikely \citep{nobles2005misleading}: since the probability of SIDS was 1 in 8543, then he claimed that the chance of having two infants dying from SIDS was
\[
\frac{1}{8543}~\times~ \frac{1}{8543} ~\approx~ \frac{1}{73,000,000}
\]
This claim was \textit{incorrect}, first of all because the product rule only works for independent events, while the deaths of two children are clearly not independent because of potential genetic and environmental factors. Moreover, the probability of the second SIDS case happening given the first case can be higher than 1 in 8543 \citep{adam2011essential}. Furthermore, Meadow's reasoning had an impact not only on the media, but also misled the jury into believing that the probability of Sally Clark being innocent was 1 in 73 million (``prosecutors' fallacy'') \citep{nobles2005misleading}, resulting in the wrongful conviction of Sally Clark. The Sally Clark's case was later studied by many statisticians, including \cite{dawid2002bayes} and \cite{hill2004multiple}, who both obtained results in favor of SIDS over murder.

Unfortunately, the Sally Clark's case is just one of many cases of wrongful convictions due to the misuse of statistics. Other notable cases include \textit{People v.~Collins} (1968) \citep{koehler1997one}, \textit{R v.~George} (2001) \citep{fenton2014neutral}, and \textit{R v.~de Berk} (2004) \citep{meester2006ab}. While the accurate application of statistics in each case was revealed during the appeal processes, the initial erroneous utilization of statistics left a lasting blemish on the role of statistics in courtrooms \citep{fenton2016bayes}.

\subsection{Cognitive bias}
\label{sec: Cognitive bias}
On March 11, 2004, a number of trains in Madrid, Spain, were bombed by terrorists, resulting in 191 people killed and over two thousands people injured. About two months later, the civil and immigration lawyer Brandon Mayfield was arrested as a suspect, as his fingerprints were thought to be found on a bag in Spain which contained similar detonation devices used in the terrorist attack \citep{national2009strengthening}. According to the FBI examination report, there were as many as 10 similar minutiae points (second level of detail, Section~\ref{sec: Fingerprint}) between the prints from Mayfield and the suspect. However, on May 19, the Spanish National Police successfully identified the latent print as the fingerprint from another person \citep{oig2006review} and the federal government paid Mayfield \$2 million for him being wrongly jailed \citep{lichtblau2006us}. Two years later, in March 2006, the Office of the Inspector General of the U.S. Department of Justice issued a thorough examination report on the factors contributing to the misidentification. The report stated that ``\textit{the verification [process] has been `tainted' by knowledge of the initial examiner's conclusion}'' \citep{oig2006review}, suggesting the impact of cognitive bias on the analysis of the evidence.

Following this incident, \cite{dror2006contextual} conducted an experiment using fingerprints that had previously been examined and identified by latent print experts as matching suspects. They then presented the same fingerprints to the same experts, but this time provided a context suggesting that they did not match, thereby implying that the suspects could not be identified. The majority of fingerprint experts made different judgements, contradicting their earlier identification decisions. This experiment confirmed the existence of cognitive bias when subjective judgement is involved in evidence analysis. 

Cognitive bias affects people's perceptions, attitudes, and decisions regarding the relevance, reliability, and applicability of statistical evidence in legal proceedings. Forensic science professionals may be prone to various forms of confirmation bias that, however, may be mitigated by limiting the access to unnecessary information, controlling the sequence of relevant information disclosure, utilizing multiple comparison samples instead of a single suspect exemplar, and ensuring result replication by analysts unaware of prior findings \citep{cooper2019cognitive}. 

\subsection{Other obstacles}
\label{sec: Other limitations}
Statistical evidence often includes information about uncertainties and errors, and it is crucial for the accuracy and transparency of forensic science techniques \citep{edmond2016model}. However, even if the statistical interpretations are effectively emphasized during a trial, it may still be difficult for laypeople, such as members of the jury, to fully understand the statistical explanations presented by forensic scientists \citep{hand1901historical,jackson2006nature,collins2019rethinking}. \cite{faigman1988bayes} performed an experiment in which a transcript from a real trial containing substantial statistical information was presented to 180 continuing education students. As expected, these students appeared to largely ignore most statistical explanations of the evidence presented by the expert witness. The results of this experiment confirmed that laypeople may be easy to logically mistake the statistical interpretation of evidence. The ``prosecutor's fallacy'' mentioned in Section~\ref{sec: The misuse of statistics} \citep{martire2020well} is just one of many cases in which the jury was easily misled into believing in incorrect statistical interpretations. 

\cite{tillers1988probability} focused on the obstacles in inferential processes, for example, the hypotheses regarding the organization of fact-finding procedures within certain legal systems. The challenges associated with the application of Bayes' Theorem have been extensively investigated \citep{tillers1986mapping,ball1960moment,tribe1971trial,nesson1978reasonable,nesson1985evidence,tribe1971continuing,finkelstein1970bayesian,finkelstein1970comment,fairley1973conversation}. Moreover, \cite{fienberg1996bayesian} discussed the difficulty of consistently applying the Bayesian framework to all evidence presented in court due to the complicated legal process, with a focus on the evaluation of statistical methods by judges and juries, because the way the questions of interest are framed can affect the interpretation of statistical evidence. \cite{fienberg2011bayesian} addressed the concerns of using the Bayesian framework in governmental and public policy settings due to its dependency on the choice of prior distributions. He used examples including U.S. election night forecasting, U.S. Food and Drug Administration (FDA) studies, to show that the adoption of Bayesian methodologies was widely acknowledged and should be embraced as the standard practice in public contexts.

\section{Likelihood ratio}
\label{sec: Likelihood ratio framework}
The likelihood ratio (LR) allows us to compare the probability of the evidence ($E$) under at least two exclusive \textit{hypotheses} or propositions \citep{curran2009statistics}. In legal proceedings, it is usually sufficient to consider the hypotheses from the prosecution view ($H_{\rm p}$) and the defense view ($H_{\rm d}$), which depend on the specific circumstances of the case, the observations made, the available background data, and the scientists' area of expertise. 
The pairs of hypotheses $(H_{\rm p},H_{\rm d})$ can be categorized into three hierarchical levels: \textit{offense}, \textit{activity}, and \textit{source}. The jury would usually address the offense level propositions, which are referred to the criminal incident itself. The forensic scientists would be more interested in the activity level and the source level, which respectively refer to the action involved in the incident and the physical evidence \citep{adam2011essential,cook1998hierarchy}. Examples of different levels of competing hypotheses are in Table~\ref{tab: hierarchy of propositions}.

\begin{table*}[htpb]
\centering
{\small
\begin{tabular}{ccll}
 \textbf{Level} & & \textbf{Homicide} & \textbf{Bank robbery} \\\hline
 \textit{Offense} & $H_{\rm p}$:&  Mr. A murdered Mr. X & Mr. S robbed the bank\\
 & $H_{\rm d}$:&  Another person murdered Mr. X & Another person robbed the bank \\\hline
 \textit{Activity} & $H_{\rm p}$:&  Mr. A stabbed Mr. X & Mr. S walked into the bank\\
 & $H_{\rm d}$:&  Mr. A was not present when Mr. X was stabbed & Mr. S was not present at the bank\\\hline
 \textit{Source} & $H_{\rm p}$:&  The blood on Mr. A's hand came from Mr. X & Mr. S left the shoe mark at the bank\\
 & $H_{\rm d}$:&  The blood on Mr. A's hand came from another person & Someone else left the shoe mark at the bank\\\hline
\end{tabular}}
\caption{\small\label{tab:widgets} Example of hierarchy of hypotheses in a homicide case \citep{cook1998hierarchy} and a bank robbery case.}
\label{tab: hierarchy of propositions}
\end{table*}

The rest of this section is organized as follows. In Section~\ref{sec: Mathematical definition}, we provide a probabilistic definition of LR. In Section~\ref{sec: Verbal interpretation}, we discuss the evolution and limitations of the verbal interpretation of LR. Finally, in Section~\ref{sec: The use in DNA typing}, we discuss the challenges of computing LR for various types of evidence, in particular DNA samples.

\subsection{Probabilistic definition}
\label{sec: Mathematical definition}
Let $A$ and $B$ be two \textit{events}. The \textit{conditional probability of $A$ given $B$} is defined as the ratio of the their joint probability and the probability of $B$, i.e.~$\mathbb{P}(A\mid B)=\mathbb{P}(A,B)/\mathbb{P}(B)$, provided that $\mathbb{P}(B)>0$. By Bayes' Theorem, this conditional probability can be rewritten as
\begin{equation}
    \mathbb{P}(A\mid B)=\frac{\mathbb{P}(B\mid A)\mathbb{P}(A)}{\mathbb{P}(B)},
\end{equation}
where $\mathbb{P}(B\mid A)$ is the conditional probability of $B$ given $A$. In the context of legal proceedings, we can use Bayes' Theorem to connect evidence $E$ with the competing hypotheses $H_{\rm p}$ and $H_{\rm d}$:
\begin{equation}
    \mathbb{P}(H_{\rm p}\mid E)=\frac{\mathbb{P}(E\mid H_{\rm p})\mathbb{P}(H_{\rm p})}{\mathbb{P}(E)}
\end{equation}
and
\begin{equation}
\mathbb{P}(H_{\rm d}\mid E)=\frac{\mathbb{P}(E\mid H_{\rm d})\mathbb{P}(H_{\rm d})}{\mathbb{P}(E)}
\end{equation}
The ratio of these two conditional probabilities yields the following fundamental factorization of the \textit{posterior odds}
\begin{equation}
    \frac{\mathbb{P}(H_{\rm p}\mid E)}{\mathbb{P}(H_{\rm d}\mid E)}=\frac{\mathbb{P}(E\mid H_{\rm p})}{\mathbb{P}(E\mid H_{\rm d})}\times\frac{\mathbb{P}(H_{\rm p})}{\mathbb{P}(H_{\rm d})},
\end{equation}
where $\mathbb{P}(H_{\rm p})/\mathbb{P}(H_{\rm d})$ is the \textit{prior odds}, which compares the prior beliefs on a suspect being guilty, and 
\begin{equation}\label{eq:lr}
    \text{LR}=\frac{\mathbb{P}(E\mid H_{\rm p})}{\mathbb{P}(E\mid H_{\rm d})},
\end{equation}
is the \textit{likelihood ratio}, or Bayes factor (see \cite{kass1995bayes}), which can be interpreted as an updating factor on prior beliefs based on the evidence $E$. When LR is smaller than 1, the posterior odds is smaller than the prior odds, i.e.~a LR smaller than 1 is in favor of the defendant. Conversely, a LR larger than 1 is in favor of the prosecution. Let us consider a simple fictional case of a bank robbery, which is an adaptation of a case presented in \citep{adam2011essential}.

\paragraph{Example: ``Bank robbery''} A shoe mark is found at a bank that has been robbed. The suspect Mr.~S is known to own a pair of shoes that can leave the same mark, and he only wears that pair of shoes. In this case, the competing hypotheses at ``source level'' (Table~\ref{tab: hierarchy of propositions}) and the evidence could be defined as
\begin{itemize}
    \item[] 
$H_{\rm p}$: Mr.~S left the shoe mark at the bank.
\vspace{-2mm}
\item[] $H_{\rm d}$: Someone else left the shoe mark at the bank.
\vspace{-2mm}
\item[] ~~$E$: The shoe mark at the bank.
\end{itemize}

\noindent To compute LR (Equation~\ref{eq:lr}) in this case, first note that $P(E\mid H_{\rm p})=1$, because if Mr.~S was the person who left the mark at the crime scene ($H_{\rm p}$), then it would match the evidence ($E$) since he only wears shoes that leave the observed mark. To compute $P(E\mid H_{\rm d})$, we need some more reasoning. Suppose that the shoes are rare and owned by only 1\% of a total of $N$ males, including Mr.~S, in a reasonably circumscribed region enclosing the crime scene. Then, we may conclude that
\begin{equation}\label{eq:pEHd}
    \mathbb{P}(E\mid H_{\rm d})=\frac{0.01\times N-1}{N}
\end{equation}
Assuming $N$ is large, we obtain $\mathbb{P}(E\mid H_{\rm d})\approx 0.01$, so LR~$ \approx 1/0.01=100$. If instead the shoes were very common and, for example, about 90\% of the $N$ males owned them, then $\mathbb{P}(E\mid H_{\rm d})\approx 0.9$, so LR~$ \approx 1/0.9=1.1$. 

In the example above, although LR is larger than 1 in both scenarios, i.e.~the evidence is in favor of the prosecution, we can definitely say that in the first case the evidence in favor of the prosecution is stronger than in the second case. However, how large should LR be to convince us that the suspect is guilty? The following section discusses this delicate challenge. 

\subsection{Verbal interpretation}
\label{sec: Verbal interpretation}
The interpretation of a LR can be vague and confusing, especially for laypeople. For example, if LR~$=100$, then this might be erroneously interpreted as ``the suspect is 100 times more possible of being guilty compared to being innocent.'' The verbal interpretation of LR has evolved dramatically over time. The first verbal scale was proposed by \cite{evett1987bayesian} (Table~\ref{tab: verbal scale}(a)), and was similar to the one proposed by \cite{jeffreys1939theory}. Then, with the rapid development of DNA typing, much larger LR values became more typically observed, so more conservative scales were proposed in \cite{evett1998interpreting} (Table~\ref{tab: verbal scale}(b)) and later in \cite{evett2000impact} (Table~\ref{tab: verbal scale}(c)). In the following years, other verbal scales were proposed, all generally considering LR~$>10^6$ as very strong or extremely strong evidence in favor of the prosecutor \citep{martire2014interpretation, marquis2016discussion}.

However, there is an ongoing debate regarding the interpretation and reporting of LRs in legal proceedings. While ordinal scales linking predefined verbal expressions to specific ranges of LR values are commonly used (\cite{champod2000commentary,providers2009standards,nordgaard2012scale,bunch2013application,willis2019enfsi}), \cite{morrison2016should} argued that LRs should be treated as estimates of true but unknown values, because derived from samples and subject to sampling uncertainty (see also \cite{beecham2011confidence}). It is therefore standard practice to report the precision of such estimates. This perspective challenges the notion that a forensic practitioner's LR reflects personal belief and has no true values to be estimated \citep{taroni2016dismissal}. Consequently, proponents of this view advocate against using verbal expressions corresponding to predetermined ranges of LR values, which are influenced by subjective human interpretation.

\begin{table*}[t]
\centering

{\footnotesize 
\begin{tabular}{c  l}
\textbf{ Likelihood ratio} &\qquad \textbf{Verbal interpretation} \\\hline
\textbf{(a)}  \hfill~~~~~~~~$1<\text{LR}\leq 10^{1/2}$~~~~~~~ &\qquad $E$ slightly increases the support for $H_{\rm p}$ against $H_{\rm d}$ \\
 $10^{1/2}<\text{LR}\leq 10^{3/2}$ &\qquad $E$ increases the support for $H_{\rm p}$ against $H_{\rm d}$ \\
 $10^{3/2}<\text{LR}\leq 10^{5/2}$ &\qquad $E$ greatly increases the support for $H_{\rm p}$ against $H_{\rm d}$ \\
~~ $10^{5/2}<\text{LR}$~~~~~~~~~~~~~~~~&\qquad $E$ very greatly increases the support for $H_{\rm p}$ against $H_{\rm d}$ \\\hline \\\hline

\textbf{(b)} \hfill $1<\text{LR}\leq 10$~~~~~~~~~~~~~~&\qquad $E$ has limited support for $H_{\rm p}$ against $H_{\rm d}$ \\
 $10<\text{LR}\leq 100$ &\qquad $E$ has moderate support for $H_{\rm p}$ against $H_{\rm d}$ \\
 $100<\text{LR}\leq 1000$ &\qquad $E$ has strong support for $H_{\rm p}$ against $H_{\rm d}$ \\
 ~~~$1000<\text{LR}$~~~~~~~~~~~~~~~~~&\qquad $E$ has very strong support for $H_{\rm p}$ against $H_{\rm d}$ \\\hline \\\hline

\textbf{(c)} \hfill  $1<\text{LR}\leq 10$ ~~~~~~~~~~~~ &\qquad $E$ has limited support for $H_{\rm p}$ against $H_{\rm d}$ \\
 $10<\text{LR}\leq 100$ &\qquad $E$ has moderate support for $H_{\rm p}$ against $H_{\rm d}$ \\
 $100<\text{LR}\leq 1000$ &\qquad $E$ has moderately strong support for $H_{\rm p}$ against $H_{\rm d}$ \\
 $1000<\text{LR}\leq 10000$ &\qquad $E$ has strong support for $H_{\rm p}$ against $H_{\rm d}$ \\
 ~~$10000<\text{LR}$~~~~~~~~~~~~~~~~~~&\qquad $E$ has very strong support for $H_{\rm p}$ against $H_{\rm d}$\\\hline
\end{tabular}}

\caption{\small Verbal interpretation likelihood ratio scales by (a) \cite{evett1987bayesian}, (b) \cite{evett1998interpreting}, and (c) \cite{evett2000impact}.}
\label{tab: verbal scale}
\end{table*}

\subsection{Computing likelihood ratios: DNA typing as an example}
\label{sec: The use in DNA typing}
Computing LR requires understanding the evidence generative process under the two competing hypotheses $H_{\rm p}$ and $H_{\rm d}$, which may be far more difficult than the example presented in Section~\ref{sec: Mathematical definition}. For pattern evidence, e.g.~fingerprints, work is in progress to formulate likelihood ratios, but substantial challenges persist. For trace evidence (e.g. glass fragments), peer-reviewed literature demonstrates the functionality of likelihood ratios, but widespread practical adoption has not occurred \citep{stern2017statistical}. However, for DNA evidence, LR is calculated and acknowledged in numerous typical situations, which we describe in the following. We start from the simplest case where the DNA samples come from only one contributor (Section~\ref{sec: Single contributor stains}), and compute LR under various assumptions of independence, population equilibria, and co-ancestry in the allele frequencies. Then, we discuss models for the more complex case of DNA mixtures, where samples come from an unknown number of individuals and proportions of contribution (Section~\ref{sec: DNA mixture likelihood framework}). We further discuss DNA typing via graphical models in Section~\ref{sec: Forensic genetics}.

\subsubsection{Single contributor samples}
\label{sec: Single contributor stains}
Suppose we have a DNA sample that originated solely from one individual, free from \textit{allelic dropout} or \textit{contamination}. The two competing hypotheses at the source level could be defined as
\begin{itemize}
    \item[$H_{\rm p}$:] The suspect is the sole contributor to the sample.\vspace{-2mm}
    \item[$H_{\rm d}$:] Someone else is the sole contributor to the sample.
\end{itemize}
We can note that $\mathbb{P}(E\mid H_{\rm p})=1$, because if the suspect is the sole contributor, then the sample from the crime scene will surely \textit{match} the suspect's genotype. To compute $\mathbb{P}(E\mid H_{\rm d})$, i.e.~the probability of match given the defendant hypothesis $H_{\rm d}$, we need to identify the \textit{relevant population}, as the estimation of the probabilities is based on the allele frequencies observed in the subpopulation that could potentially be a match besides the suspect \citep{buckleton2005forensic,national1996evaluation,balding2015weight}. Most forensic labs identify the relevant population by relying on their own databases or data published by agencies, such as the CODIS provided by the FBI \citep{budowle1999population,budowle2001codis,moretti2016population,FBICODIS}, and then compute probabilities by making some assumptions.

\paragraph{Independence assumptions.}
The simplest assumption for computing genotype probabilities in the relevant population is the independence between alleles \citep{buckleton2005forensic,crow2017introduction}. In this case, the population is considered to be in Hardy-Weinberg equilibrium (HWE) where, because of independence between alleles, the genotype probabilities can be computed as the product of allele probabilities \citep{hardy1908mendelian, weinberg1908vererbungsgesetze}: if $A_{i\text{1}}$ and $A_{i\text{2}}$ are two alleles of genotype $G$ at locus $i$, say $G_i$, then the genotype probability is given by
\begin{equation}
    \mathbb{P}(G_i)=
    \begin{cases}
        \ \mathbb{P}^2(A_{i\text{1}})&, A_{i\text{1}}=A_{i\text{2}}\\
        \ 2\mathbb{P}(A_{i\text{1}})\mathbb{P}(A_{i\text{2}})&, A_{i\text{1}}\neq A_{i\text{2}}
    \end{cases}
\end{equation}
Furthermore, alleles at different loci are considered independent if the population is in linkage equilibrium (LE) \citep{bennett1952theory}, so the multi-loci genotype probability is given by
\begin{equation}
    \mathbb{P}(G)=\prod_i\mathbb{P}(G_i).
\end{equation}
Therefore, with estimates of the relevant allele probabilities, we can then obtain, as in Equation~\ref{eq:pEHd}, $P(E\mid H_d)=\frac{\mathbb{P}(G)N-1}{N}\approx\mathbb{P}(G)$ for large $N$, so LR$\approx 1/\mathbb{P}(G)$. The necessary conditions for HWE that would make sense of the independence assumptions are, unfortunately, unrealistic: (i) population size is infinitely large; (ii) entirely random mating, including selfing; (iii) absence of disruptive factors such as migration, mutation, or selection. Indeed, the independence assumptions discussed above raised concerns about overestimating the weight of DNA evidence \citep{curran2009statistics}, as in a small or inbred population, the probability of some other person having the same genotype is higher than in HWE, yielding a larger $\mathbb{P}(E\mid H_{\rm d})$ and thereby a smaller LR.

\paragraph{Co-ancestry assumption.}
\label{sec: Co-ancestry assumption}
\cite{wright1949genetical} proposed the \textit{co-ancestry coefficient}, here denoted by $\theta\in [0,1]$. This coefficient measures the average progress of subpopulations towards \textit{fixation} \citep{balding2015weight}, which is the evolutionary state where all individuals within a specific subpopulation share the same genotype, i.e.~for a single locus, only one type of allele can be observed. Thus, $\theta$ can be interpreted as the ``proportion of alleles that share a common ancestor in the same subpopulation'' \citep{balding1994dna}, with $\theta=1$ corresponding to fixation, and $\theta=0$ denoting a homogeneous population, where the proportions of alleles are identical in every subpopulation. \cite{nichols1991effects} and \cite{balding1994dna} used the co-ancestry coefficient as a parameter of a Beta-Binomial distribution \citep{curran2009statistics} to model allele frequencies. Suppose we observe $n$ allelic samples, and for the $i$-th sample let $X_i=1$ if allele $\omega$ is present and $X_i=0$ otherwise. Then, assume
\begin{eqnarray}
    X_1,...,X_n \mid p_\omega \hspace{-2mm}&\stackrel{\rm i.i.d.}{\sim}&\hspace{-2mm} {\rm Bernoulli}(p_\omega),\\
    p_\omega \hspace{-2mm}&\sim&\hspace{-2mm} {\rm Beta}\left((\theta^{-1}-1)\bar{p}_\omega,~(\theta^{-1}-1)(1-\bar{p}_\omega)\right).~~~~~~~~~
\end{eqnarray}
where $\theta\in [0,1]$ is the co-ancetry parameter, $\bar p_\omega\in [0,1]$ is the relative frequency of allele $\omega$ in the full population, and $p_\omega\in[0,1]$ represents the (random) relative frequency of allele $\omega$ at a given locus in the relevant subpopulation of interest. Note that $\mathbb{E}[p_\omega]=\bar p_\omega$ and ${\rm Var}(p_\omega)=\bar p_\omega(1-\bar p_\omega)\theta$, and that $X_1,...,X_n$ are independent conditionally on $p_\omega$, but not unconditionally. Moreover, if $\theta=1$ (fixation), $p_\omega$ will follow a Bernoulli distribution across all of the subpopulations, as it can either take value of 0 or 1, with parameter $\bar{p}_\omega$; whereas if $\theta=0$ (homogeneity), the allele frequencies across all subpopulations will be the same, thus $p_\omega$ will be a constant and the variance will be 0. In order to compute probabilities from the Beta-Binomial model efficiently, let us define the function
\begin{eqnarray}
q_\omega(m,n)\hspace{-2mm}&:=&\hspace{-2mm}\mathbb{P}(\omega\mid \text{observed}~m~\text{alleles of type}~\omega~\text{out of}~n)~~~~~~~\\
\hspace{-2mm}&~=&\hspace{-2mm}\mathbb{P}\left(X_{n+1}=1\mid \sum_{i=1}^n X_i=m\right) = 
\frac{m\theta+\bar p_\omega(1-\theta)}{1+(n-1)\theta}
\end{eqnarray}
Suppose that the suspect genotype at locus $i$ is $G_{\rm S_i}=AA$ (homozygous), where $A$ is the allele type modeled by the Beta-Binomial distribution with parameters $\theta$ and $\bar p_A$. Then, the probability of the genotype at locus $i$ of another person (the culprit $G_{\text{C}_i}$) in the subpopulation being the same as the suspect is
\begin{equation}\label{eq: pAA}
    \mathbb{P}(G_{\text{C}_i}\mid G_{\text{S}_i})=        \mathbb{P}(AA\mid AA)  =q_A(2,2)q_A(3,3) 
\end{equation}
In the case where the suspect genotype at locus $i$ is $G_{{\rm S}_i}=AB$ (heterozygous), where $B$ is another allele type also modeled by the Beta-Binomial distribution with parameters $\theta$ and $\bar p_B$, then we would obtain
\begin{equation}\label{eq: pAB}
    \mathbb{P}(G_{\text{C}_i}\mid G_{\text{S}_i})= 
        \mathbb{P}(AB\mid AB)= q_A(1,2)q_B(1,3)+q_B(1,2)q_A(1,3) 
\end{equation}
Equations~\ref{eq: pAA} and~\ref{eq: pAB} can be used to compute the denominator of LR. For example, under the assumption of linkage equilibrium (LE) \citep{bennett1952theory}, the multi-loci genotype probability is given by $\mathbb{P}(G_{\rm C}\mid G_{\rm S})=\prod_i \mathbb{P}(G_{\text{C}_i}\mid G_{\text{S}_i})$, and therefore, $\text{LR}=1/\mathbb{P}(G_{\rm C}\mid G_{\rm S})$.

The parameter $\bar p_\omega$ can be estimated empirically from nation-wise databases. Efforts were made to estimate the co-ancestry coefficient $\theta$: method of moments estimation \citep{weir1984estimating,weir1996genetic,weir2002estimating}, likelihood-based inference \citep{balding2003likelihood}, and Bayesian methods \citep{balding1996population} can be used when relevant population data is available. At conventional loci, estimated values are often much less than 1\% \citep{morton1992genetic}. Otherwise, if the relevant data is unavailable or too scarce to yield reliable estimates, $\theta$ is often set equal to some biologically justified upper bound, e.g. if $\theta=0.0625$, individuals in the subpopulation are at least related as first cousins, and if $\theta=0.25$, all individuals are related as siblings \citep{curran2009statistics}.

\subsubsection{DNA mixtures}
\label{sec: DNA mixture likelihood framework}
In some cases, especially rape and sexual assault, DNA samples are \textit{mixed} from multiple contributors, where we usually detect more than two alleles at one locus. However, even in cases where samples appear to have originated from a single contributor and exhibit only one or two alleles per locus, they could potentially be mixtures where one contributor is concealed by another one (masking effect; \cite{buckleton2007towards}). Thus, it is usually impossible to identify the number of contributors in a DNA sample correctly. \cite{paoletti2005empirical} showed that around 3\% of mixtures involving three individuals would be incorrectly identified as mixtures of only two individuals, and over 70\% of mixtures involving four individuals would be inaccurately classified as mixtures of two or three individuals if only the maximum number of alleles observed at any tested locus is considered.

Peak height or area information from the electropherogram (Section~\ref{sec: DNA}) can be used in analyzing mixed DNA samples, as the major contributor will make larger peaks \citep{buckleton2007towards}: if two DNA profiles are mixed together, the approximate ratio will remain consistent when comparing the peak areas or heights of different component alleles within a locus \citep{lygo1994validation,sparkes1996validation}. Yet, the computational complexity increases very quickly with the number of contributors \citep{clayton1998analysis}.

Typically, examiners utilize the profiles of individuals who can be reliably assumed to be present in the mixture, and consistently incorporate the genotype of the profile of interest, for example, the suspect \citep{buckleton2008discussion}. Because of the complexity due to unclear number of contributors, most laboratories assign the number based on the minimum number required to account for the observed peaks. However, adopting such an approach may result in non-exhaustive hypotheses, as they fail to include scenarios involving a greater number of contributors \citep{buckleton2007towards}. The computation of LR has been extensively studied in these scenarios by 
\cite{gill2006dna,weir1997interpreting,curran1999interpreting,gill1998interpreting,evett1998interpreting,evett1991guide,clayton1998analysis,bill2005pendulum}.

Another approach to the analysis of mixed DNA samples is the ``random man not excluded'' method (RMNE), which aims to ascertain the proportion of the population that could potentially be included as contributors to the stain found at the crime scene. RMNE does not make any assumptions regarding the number of contributors, which is widely regarded as a significant advantage. However, RMNE has also garnered substantial criticism and negative feedback \citep{buckleton2007towards,national1996evaluation,weir1999court}. \cite{brenner1997s} used a paternity case as an example to demonstrate that RMNE may waste information and considered it as ``dishonest and unsatisfactory.''

\begin{figure*}[t!]
    \centering
    \subfigure[Simple BN.]{
        \begin{minipage}{7.5cm}
            \centering
            \includegraphics[width=1\textwidth]{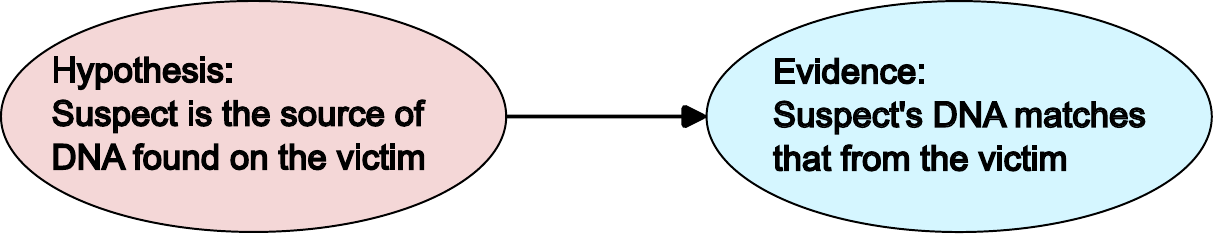}
        \end{minipage}
    }
    \subfigure[More complex BN.]{
        \begin{minipage}{7.5cm}
            \centering
            \includegraphics[width=.82\textwidth]{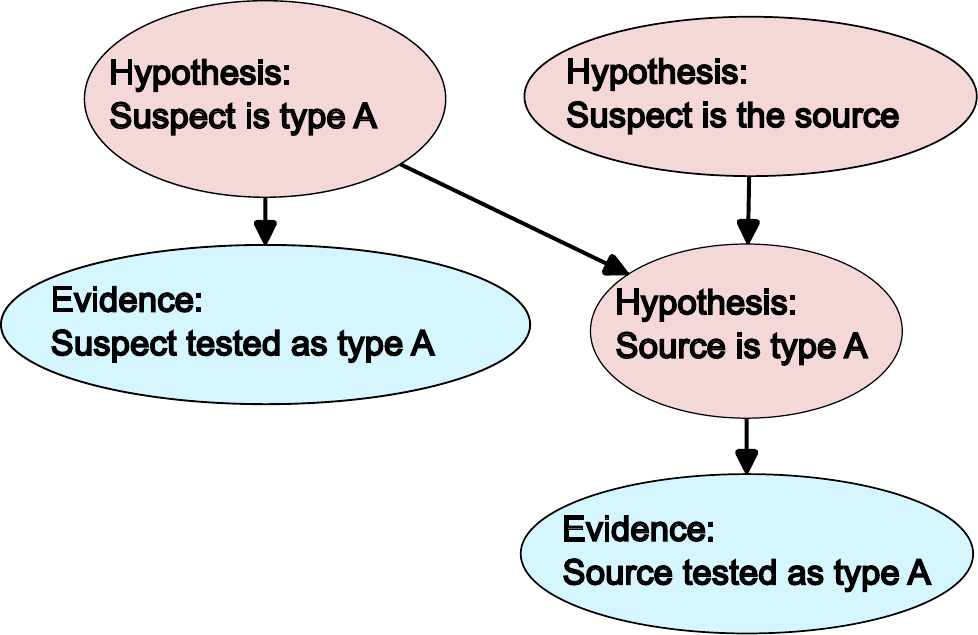}
        \end{minipage}
    }
    \caption{\small\textbf{BNs for the ``Sexual assault'' example.} \textbf{(a)} Simple BN with a pink node representing the prosecutor's hypothesis $H_{\rm p}$, and a blue node representing the evidence $E$. The arrow signifies a causal link in which the observed evidence depends on whether the hypothesis is true, but not the other way around. \textbf{(b)} More complex causal structure for the same case, where the prosecutor's hypothesis is now decomposed into three hypotheses (pink nodes) and the evidence is decomposed into two components (blue nodes). If the suspect's DNA is type A, then the suspect's DNA \textit{should} be tested as type A. If the suspect's DNA is type A and \textit{also} the source of the DNA sample found on the victim, then the source must be type A. Finally, if the source is type A, then the DNA sample found on the victim \textit{should} be tested as type A.}
    \label{fig: BN DNA}
\end{figure*} 

\section{Graphical models}
\label{sec: Graphical models}
Forensic investigations often involve multiple types of evidence that are related to each other, and it is important to take these dependencies into account while assessing competing hypotheses. \textit{Bayesian Networks} (BN) are probabilistic graphical models that allow us to describe forensic investigations via \textit{directed acyclic graphs} (DAG), where \textit{nodes} or vertices represent hypotheses and evidence, and \textit{arrows} connect nodes to describe statistical dependence and causality. BNs allow us to compute conditional probabilities and likelihood ratios involving multiple types of evidence simultaneously. Different types of graphs can be used to represent hypotheses and evidence with different levels of detail. Let us first consider a simple example, which is an adaptation of a case presented in \cite{fenton2016bayes}.\vspace{-2mm}
\paragraph{Example: ``Sexual assault''} DNA is found on the victim of a sexual assault. A suspect is arrested. We have
\begin{enumerate}
    \item[$H_{\rm p}$:] The suspect is the source of the DNA found on the victim. \vspace{-2mm}
    \item[$H_{\rm d}$:] Someone else is the source of the DNA found on the victim.
    \vspace{-6mm}
    \item[$E$:] The suspect's DNA matches the one found on the victim.
\end{enumerate}
In Figure~\ref{fig: BN DNA}(a) we show a simple graph with a pink node representing the prosecutor's hypothesis $H_{\rm p}$, and a blue node representing the evidence $E$. The arrow signifies a causal link in which the observed evidence depends on whether the hypothesis is true, but not the other way around. Figure~\ref{fig: BN DNA}(b) depicts a more complete causal structure for the same case, where the prosecutor's hypothesis is now decomposed into three hypotheses ``suspect's DNA is type A'', ``suspect is the source'', and ``source is type A'', and the evidence is decomposed into two components ``suspect is tested as type A'' and ``source is tested as type A'' \citep{dawid1998forensic,fenton2014calculating}. The arrows in this graph can be interpreted as follows. If the suspect's DNA is type A, then the suspect's DNA \textit{should} be tested as type A. If the suspect's DNA is type A and \textit{also} the source of the DNA sample found on the victim, then the source must be type A. Finally, if the source is type A, then the DNA sample found on the victim \textit{should} be tested as type A. This second graph let us better analyse potential errors from DNA collection and testing process \citep{foreman2003interpreting, koehler1993error, thompson2003probability}. This graph may also be expanded further by including additional nodes such as the hypothesis ``the suspect was at the scene'' to better evaluate the likelihood of the observed evidence.

This section is organized as follows. In Section~\ref{sec: Object-oriented networks}, we present a special class of BNs, the object-oriented BNs, which are graphs with a versatile hierarchical structure where nodes can be expanded into sub-graphs to enhance interpretation and analysis of the various hypotheses and evidence components in a forensic investigation. In Section~\ref{sec: Conditional independence}, we discuss strategies for the interpretation and probability computation in BNs. In Section~\ref{sec: Forensic genetics}, we illustrate the application of BNs in DNA typing. Finally, in Section~\ref{sec: Chain event graph (CEG)}, we briefly describe chain event graphs, another type of graphical models.

\subsection{Object-oriented Bayesian Networks}
\label{sec: Object-oriented networks}
Object-oriented Bayesian Networks (OOBNs) are BNs where some nodes called \textit{modules} may further represent graphs. The hierarchical structure of OOBNs enhances the comprehensibility of the graph and facilitates modifications based on various scenarios and constraints \citep{dawid2018graphical,laskey2013network}. Let us consider an example, which is an adaptation of a fictional crime presented in \cite{dawid2018graphical}.

\paragraph{Example: ``Store burglary''} A man breaks into a closed store by smashing a window, and then punches the cashier in the face before stealing the money. Based on the information provided by the cashier, a suspect is arrested and the forensic team collects and analyses glass and blood samples found on his clothing. The glass sample is indistinguishable from the glass of the broken window, and the DNA in the blood sample matches the cashier's DNA. 

\begin{figure*}[t]
    \centering
        \begin{minipage}{7cm}
            \centering
     \subfigure{
           \includegraphics[width=.6\textwidth]{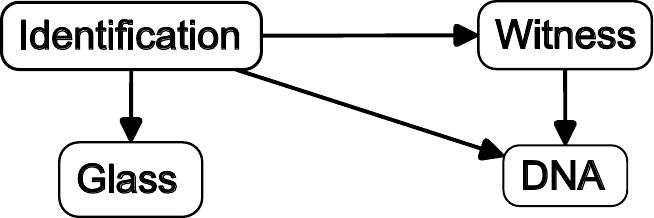}}\\{\footnotesize(a) High-level OOBN.}\\~\\
             \subfigure{
            \includegraphics[width=1\textwidth]{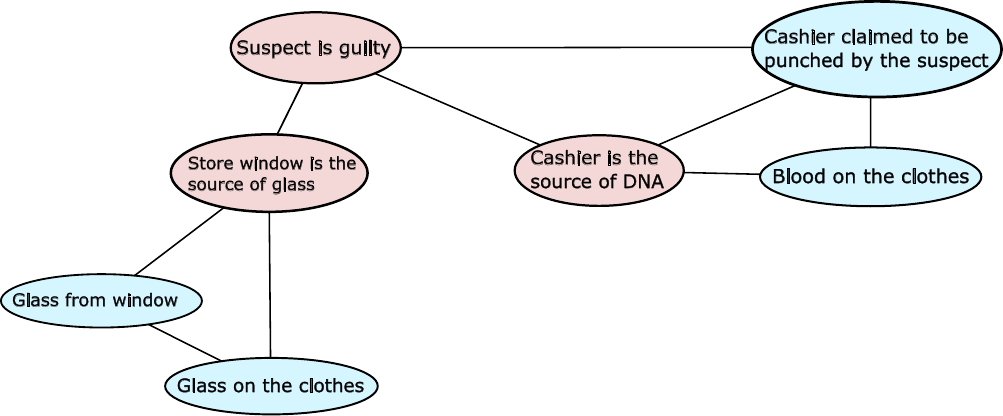}
            }\\{\footnotesize(c) Moralised graph.}
        \end{minipage}~~
    \subfigure{
        \begin{minipage}{11cm}
            \centering
            \includegraphics[width=.97\textwidth]{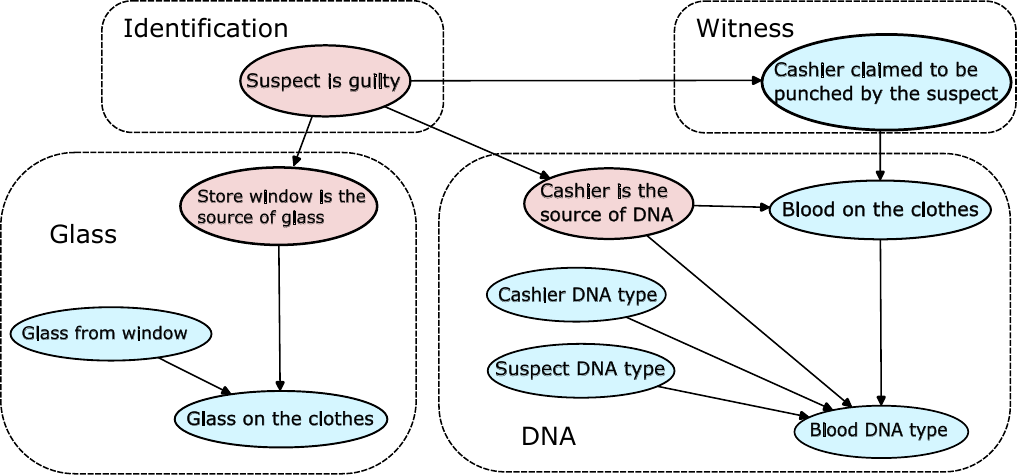}\\{\footnotesize(b) Expanded OOBN.}
        \end{minipage}
    }
    \caption{\small\textbf{OOBNs for the ``Store burglary'' example.} \textbf{(a)} High-level OOBN containing the unobserved \textit{identification} module, and three observed evidence modules \textit{witness}, \textit{glass}, and \textit{DNA}. The arrows describe the dependence of the evidence on the identity of the suspect. \textbf{(b)} Expanded OOBN where each dashed region surrounds the sub-graph of the corresponding higher-level module in (a). The pink nodes represent unobserved hypotheses, while the blue nodes represent observed evidence. \textbf{(c)} Moralised graph for conditional independence analysis. This moralized graph was built with the goal of assessing the conditional dependence between ``blood on the clothes'' and ``glass on the clothes'' in the OOBN in (b). Since all undirected paths connecting the two nodes of interest pass through ``suspect is guilty'', we can conclude that ``blood on the clothes'' and ``glass on the clothes'' are independent conditionally on knowing the guilt or innocence of the suspect.}
    \label{fig: OOBN simple example}
\end{figure*}

In this case, the goal is the identification of the suspect based on the three evidence components: witness, glass, and DNA. The graph shown in Figure~\ref{fig: OOBN simple example}(a) is a \textit{high-level OOBN} containing four modules: the unobserved \textit{identification} module, and three observed evidence modules \textit{witness}, \textit{glass}, and \textit{DNA}. The arrows describe the dependence of the evidence on the identity of the suspect. Each module can be expanded to a BN, yielding the expanded BN shown in Figure~\ref{fig: OOBN simple example}(b). Each dashed region surrounds the sub-graph of the corresponding higher-level module in Figure~\ref{fig: OOBN simple example}(a). The pink nodes represent unobserved (Boolean) hypotheses, while the blue nodes represent observed evidence. The \textit{identification} module includes only one node, which represents the prosecutor's hypothesis ``the suspect is guilty.'' Arrows departing from this node point to the hypotheses about the evidence ``the store window is the source of glass retrieved from the suspect's clothes'' and ``the cashier is the source of the DNA retrieved from the blood on the suspect's clothes'', as well as the witness testimony about ``being punched by the suspect.'' If the testimony is trustworthy and the cashier is indeed the source of the DNA, we would expect to see blood on the suspect's clothes. To analyze the DNA samples obtained from the blood on the suspect's clothes, we need the blood sample, DNA profile from both cashier and suspect, and the hypothesis on cashier being the source of the DNA. Similarly, if ``the store window is the source of glass'' is true, then together with the evidence ``glass from the store window,'' which is collected by the investigators after the crime, we should expect a match between the two glass samples. 

OOBNs are very versatile models that can be easily expanded by incorporating additional features \citep{dawid2007object}. This is especially possible with the use of \textit{generic modules}, i.e.~modules with pre-defined structures that can be employed in novel combinations, both within and across higher-level networks \citep{neil2000building,hepler2007object,fenton2013general}. Numerous generic modules have been developed for different objects, including testimony \citep{dawid2011inference}, identification, contradiction, corroboration, conflict, and convergence \citep{hepler2007object}. In Section~\ref{sec: Forensic genetics}, we illustrate the use of the generic modules identification, child, and founder for DNA typing.

\subsection{Interpretation and computation of probabilities in Bayesian Networks}
\label{sec: Conditional independence}
Probabilistic graphical models offer several advantages, including a concise visual illustration of the factorization of the joint probability distribution of a statistical model, which is made possible due to the graphical models' ability to represent conditional independence relationships among the random variables or events \citep{dawid1979conditional,lauritzen1996graphical}. Therefore, we can make efficient inference and make it accessible to individuals without formal statistical or mathematical training, enabling collaboration among statisticians, domain experts, and decision makers \citep{walley2023cegpy}.

\subsubsection{Conditional independence and moralization}
Similar to the definition of independent events, for which the joint probability of two events $A$ and $B$ factorizes into the product of their marginal probabilities, i.e.~$\mathbb{P}(A,B)=\mathbb{P}(A)\cdot \mathbb{P}(B)$, we say that $A$ and $B$ are independent conditionally on $C$ if $\mathbb{P}(A,B\mid C)=\mathbb{P}(A\mid C)\cdot\mathbb{P}(B\mid C)$. Conditional independence can sometimes impose the irrelevance of evidence, and also substantially simplify probability calculation on the graph. In BNs, arrows connecting nodes are associated with conditional independence relationships. \cite{lauritzen1990independence} proposed the \textit{moralisation} criterion to identify conditional independence in directed graphs efficiently. To conclude the method, suppose we want to study the independence between graph nodes $S$ and $T$ conditionally on $U$. We first form the \textit{ancestral graph}, which is the subgraph containing only nodes $S$, $T$, $U$, and their \textit{ancestors}, i.e.~all parent nodes with directed \textit{paths} (i.e.~sequence of links) leading to $S$, $T$, or $U$. Then, we add undirected \textit{edges} between any disconnected pairs of parents of a common child (``marrying the unmarried parents''). Finally, we replace all directed links with undirected edges. The resulting undirected graph is called \textit{moralized graph}. If all paths that connect $S$ and $T$ pass through node $U$, then $S$ and $T$ are independent conditionally on $U$. In Figure~\ref{fig: OOBN simple example}(c) we show the moralized graph of the OOBN in Figure~\ref{fig: OOBN simple example}(b), to assess the dependence between the two nodes ``blood on the clothes'' and ``glass on the clothes'' conditionally on ``suspect is guilty.'' Since all paths connecting the two nodes of interest pass through ``suspect is guilty'', we can conclude that knowing the guilt or innocence of the suspect ensures that the evidence related to glass and DNA remains independent of each other. Similar conclusions would be drawn from the moralized graph of the OOBN in Figure~\ref{fig: OOBN simple example}(a), as the glass module and the DNA module would be found conditionally independent given the identification module. All these conclusions make sense since, qualitatively speaking, glass and DNA are clearly physically unrelated.

\subsubsection{Computing conditional probabilities}
\label{sec: estimation}
BNs offer an elegant approach to manipulating probability distributions. BNs enable the specification of conditional distributions for each node based on the states of its \textit{parent} variables, eliminating the need to specify a vast collection of joint probabilities for all variables involved in the problem \citep{dawid2018graphical}. In BNs used for forensic science, specific variables within the graphs are assigned predetermined values and/or probabilities based on previous studies, for example, the population frequency of an allele. Our objective is to determine the conditional probabilities of other variables of interest, for example, the probabilities involved in the computation of LR (Equation~\ref{eq:lr}), and often these computations can be implemented efficiently only via \textit{algorithms}.

\textit{Belief propagation} is a probability propagation method for the \textit{exact inference} of probabilities over tree-structured graphs, and it is implemented via a local message-passing algorithm \citep{pearl1982reverend,pearl1988probabilistic,lauritzen1988local}. Belief propagation treats each variable as a simple processor and examines asynchronous local message passing among different nodes until equilibrium is attained. Belief propagation can be implemented efficiently with several existing software packages in R, Python, and other platforms.

In certain scenarios, however, the exact computation of probabilities via belief propagation may be difficult or infeasible. For example, in the presence of large cliques or extended loops, the convergence of the belief propagation algorithm is not guaranteed \citep{frey1997revolution}, and if the graph contains cycles, the algorithm no longer works. In these cases, other algorithms for exact inference, such as the \textit{junction tree algorithm}, could be used \citep{madsen1999lazy,barber2004probabilistic,kahle2008junction}. Alternatively, \textit{approximate inference} methods are available, including variational methods \citep{jaakkola1997variational,jordan1999introduction} and sampling methods \citep{cheng2015efficient,feng2019dynamic}.

\subsection{Graphical models in DNA typing}
\label{sec: Forensic genetics}
Besides the likelihood ratio framework described in Section~\ref{sec: The use in DNA typing}, BNs are widely used in the analysis of forensic genetics evidence due to the nature of DNA, which is inherited from parents to children, a natural causal link used to build BNs. \cite{dawid2002probabilistic} introduced the use of BNs in the analysis of forensic genetics evidence, and \cite{dawid2007object} investigated the use of OOBNs. In Section~\ref{sec: Single contributor stains graphical models}, we present graphical models for DNA typing from single contributor samples in criminal and civil paternity cases, and discuss different assumptions on founder genes, including the modeling of allele frequencies, dependence between individuals, and the role of subpopulations. In Section~\ref{sec: DNA mixture graphical models}, we present graphical models for the analysis of mixed DNA samples.

\begin{figure*}[t]
    \centering
    \subfigure[OOBN for DNA typing in a criminal case.]{
        \begin{minipage}{10.5cm}
            \centering
            \includegraphics[width=.95\textwidth]{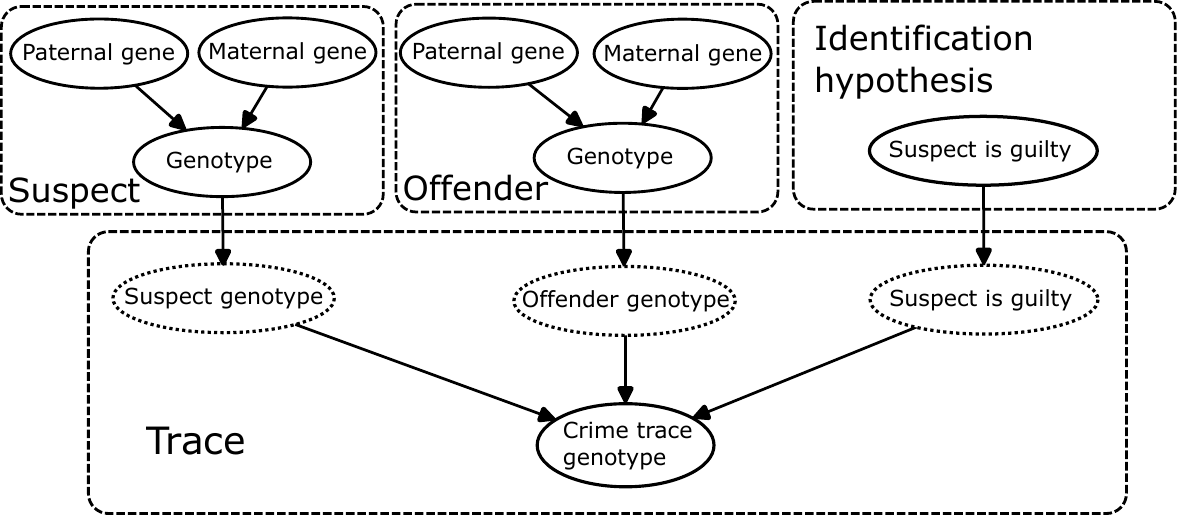}
        \end{minipage}
    }~
    \subfigure[OOBN for DNA typing in a civil paternity case.~~~~~~~~~~~~~~~]{
        \begin{minipage}{7.5cm}
            \centering
            \includegraphics[width=.95\textwidth]{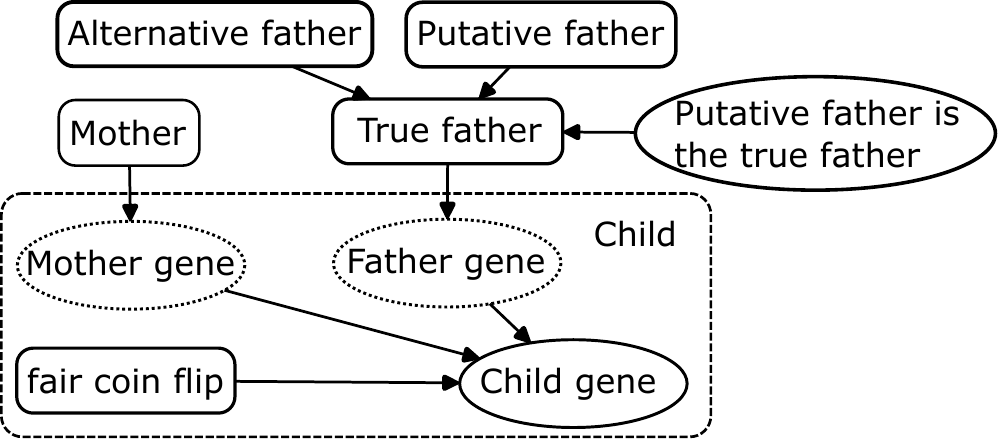}
        \end{minipage}
    }
    \caption{\small\textbf{OOBNs for DNA typing from single contributor samples}. \textbf{(a)} OOBN for the criminal case ``Sexual assault'' example. The graph contains the modules \textit{suspect}, \textit{offender}, \textit{trace}, and the identification hypothesis ``suspect is guilty.'' Both \textit{suspect} and \textit{offender} are instances of generic founder modules that include the nodes ``paternal gene'', ``maternal gene'', and their unordered combination ``genotype.'' Module \textit{trace} is an instance of generic identification module where the \textit{crime trace genotype} takes value of ``suspect genotype'' or ``offender genotype'' based on the true or false of the hypothesis respectively. \textbf{(b)} OOBN for DNA typing in a civil paternity case. The graphs contains three generic founder modules \textit{mother}, \textit{putative father}, \textit{alternative father}, a generic identification module \textit{true father}, the hypothesis ``putative father is the true father'', and a generic child module, where the child's gene is randomly drawn via fair coin flip from either the maternal gene or the paternal gene.}
    \label{fig: OOBN criminal and civil paternity}
\end{figure*}

\subsubsection{Single contributor samples}
\label{sec: Single contributor stains graphical models}
Suppose we have a sample that originated solely from one individual, free from allelic dropout or contamination.

\paragraph{Simple criminal and civil paternity cases.}

Figure~\ref{fig: OOBN criminal and civil paternity}(a) shows an OOBN with modules \textit{suspect}, \textit{offender}, \textit{trace}, and the identification hypothesis node describing the criminal case example ``Sexual assault''. Modules \textit{suspect} and \textit{offender} are generic founder modules consisting of founding genes nodes ``paternal gene'', ``maternal gene'', and their unordered combination ``genotype.'' The \textit{trace} module is an instance of the generic identification module, where the input comes from the suspect genotype and offender genotype, i.e.~the output of corresponding modules. Node ``crime trace genotype'' is identical to ``suspect genotype'' or ``offender genotype'' given the true or false of the hypothesis ``suspect is guilty'' respectively.

In simple civil paternity cases, the goal is to find the likelihood ratio of the hypothesis of paternity ``putative father is the true father'' against the non-paternity hypothesis ``the true father is an unspecified alternative father'' who is unrelated to the putative father and is drawn from a reasonable sub-population. The OOBN in Figure~\ref{fig: OOBN criminal and civil paternity}(b) contains the generic founder modules \textit{mother}, \textit{putative father}, \textit{alternative father}, the generic identification module \textit{true father}, the hypothesis ``putative father is the true father'', and a generic child module (a.k.a. meiosis module) to represent Mendelian inheritance at meiosis stage, where the ``child gene'' is randomly drawn from either the maternal gene or paternal gene denoted by the node ``fair coin flip''.

Thanks to the flexibility of OOBNs, the simple OOBNs described above can be easily modified to allow for \textit{mutations} \citep{dawid2001non,vicard2004statistical}, and to deal with more complicated paternity cases where, for example, the DNA profile of the putative father is unavailable, while it is available from one of his siblings and other relatives.

\begin{figure*}[t]
    \centering
    \subfigure[Alleles generative network (P\'{o}lya urn scheme) with uncertain allele frequencies.]{
        \begin{minipage}{3.5cm}
            \centering
            \includegraphics[width=1\textwidth]{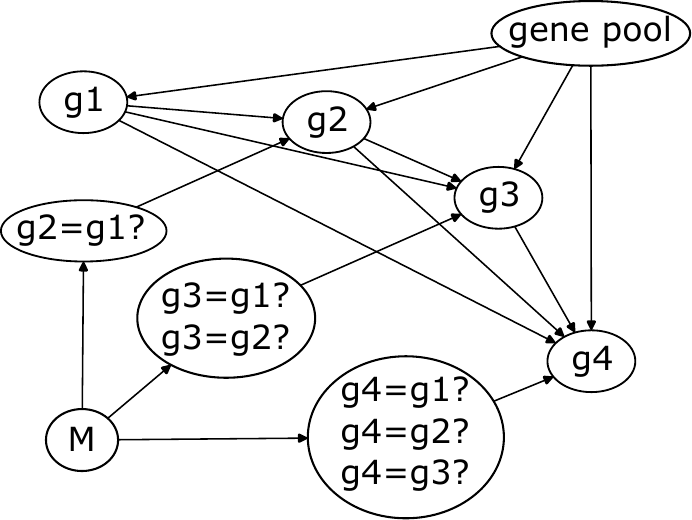}
        \end{minipage}
    }~
    \subfigure[Multi-loci DNA typing with uncertain subpopulation.]{
        \begin{minipage}{14cm}
            \centering
            \includegraphics[width=1\textwidth]{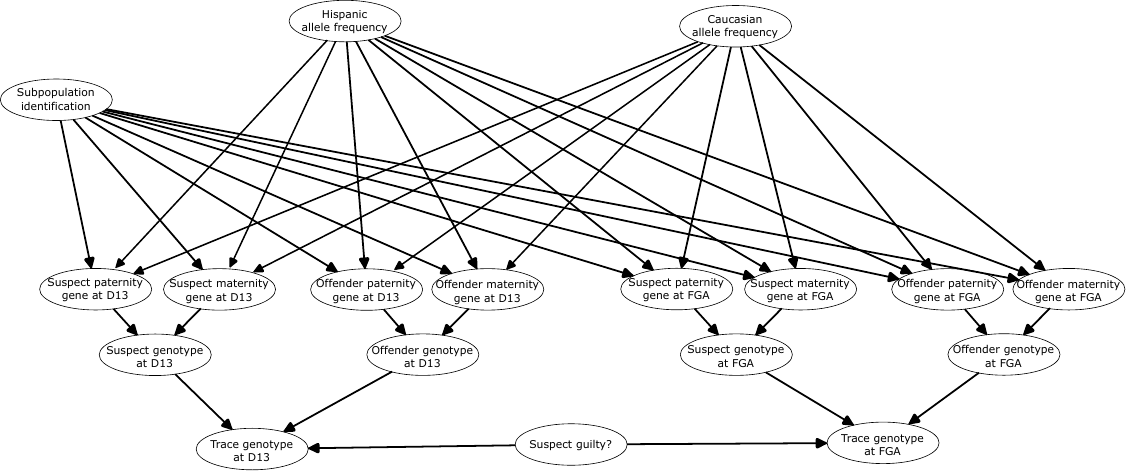}
        \end{minipage}
    }
    \caption{\small\textbf{Allele generative models}. \textbf{(a)} Alleles generative network with uncertain allele frequencies. Alleles $g_1,...,g_4$ are generated sequentially (via P\'{o}lya urn scheme) from a Multinomial model with random allele frequencies drawn from the model in Equation~\ref{eq: post}. \textbf{(b)} Multi-loci DNA typing with uncertain subpopulation. The node ``subpopulation identification'' determines whether the founder genes ``suspect paternity'', ``suspect maternity'', ``offender paternity'' and ``offender maternity'' are drawn from the Hispanic subpopulation or the Caucasian subpopulation. The node ``suspect is guilty'' determines whether the trace genotype at \texttt{D13} and \texttt{FGA} takes values from suspect/offender genotypes.}
    \label{fig: DNA typing generative}
\end{figure*}

\paragraph{Assumptions on founder genes.} 
\label{sec: Assumptions on founder genes}
The conventional assumptions of fixed and known allele frequencies, independence among individuals in the model, and homogeneity in the allele frequency databases substantially simplify the computation of conditional probabilities from BNs of DNA typing. However, these assumptions can all be subject to scrutiny \citep{green2009sensitivity}. Besides the methods introduced in Section~\ref{sec: Single contributor stains}, additional structures can be introduced in graphical models to model these complexities. Here, we present an approach to model the uncertainty about allele frequencies.

As explained in Section~\ref{sec: The use in DNA typing}, in practice, the allele relative frequencies used for probabilistic forensic inference are not known probabilities but rather estimates derived from empirical allele frequencies in a database. \cite{green2009sensitivity} proposed a Bayesian approach to model the uncertainty in allele frequencies. Suppose $G_1,\ldots,G_N$ are the alleles at one specific locus observed in a genetics database. Assuming there exist only $K$ possible allele types at this locus, let $Y_j=\sum_{i=1}^N\mathbbm{1}\{G_i=\text{type }j\}$ denote the number of observed alleles of type $j\in\{1,...,K\}$, and further assume
\begin{equation}
(Y_1,\ldots,Y_K)\sim\text{Multinomial}(N,p_1,\ldots,p_K)
\end{equation}
where $p_1,...,p_K$ are allele probabilities, $\sum_{j=1}^Kp_j=1$, and $\sum_{j=1}^KY_j=N$. Moreover, assume the prior distribution
\begin{equation}
(p_1,\ldots,p_K)\sim\text{Dirichlet}(\alpha_1,\ldots,\alpha_K),
\end{equation}
where $\alpha_1,...,\alpha_K>0$ are prior parameters. This prior induces the posterior distribution
\begin{equation}\label{eq: post}
(p_1,\ldots,p_K)\mid (Y_1,\ldots,Y_K)\sim\text{Dirichlet}(\alpha_1+Y_1,\ldots,\alpha_K+Y_K).
\end{equation}
Then, the authors propose to model the randomness of alleles, say $g_1,...,g_m$, in DNA typing cases as being drawn independently from the pool of $K$ allele types with random probabilities $\tilde p_1,...,\tilde p_K$ sampled from the posterior distribution in Equation~\ref{eq: post}. This sampling corresponds to the standard set-up of a Dirichlet process \citep{ferguson1973bayesian} and can be represented as a Bayesian network using a P\'{o}lya urn scheme \citep{blackwell1973ferguson}, where the alleles $g_1,...,g_m$ can be thought as being drawn sequentially as follows. We first draw $g_1$ from the pool of $K$ allele types with probabilities $\hat p_1,...,\hat p_K$, where $\hat p_j = \frac{\alpha_j+Y_j}{M}$ and $M=N+\sum_{j=1}^K\alpha_j$. Then, for $i=2,...,m$, $g_i$ is drawn uniformly from $g_1,\ldots,g_{i-1}$ with probability $\frac{i-1}{M+i-1}$, and with probability $\frac{M}{M+i-1}$ it is drawn from the allele pool with probabilities $\hat p_1,...,\hat p_K$. Figure~\ref{fig: DNA typing generative}(a) shows the graph of the sampling process of four founding genes $g_1 =$ suspect paternity, $g_2 =$ suspect maternity, $g_3 =$ offender paternity, and $g_4 =$ offender maternity. Graphs like this one can be used as modules of OOBNs, such as the ones presented in Figure~\ref{fig: OOBN criminal and civil paternity}.

\paragraph{Co-ancestry assumptions.}
The assumption of a homogeneous DNA reference population is also questionable as populations often comprise diverse subgroups \citep{green2009sensitivity}. The uncertainty regarding the relevant population can also induce observed or unobserved dependence between actors, as well as marker dependence in cases involving untyped individuals. Figure~\ref{fig: DNA typing generative}(b) (adapted from \cite{green2009sensitivity}) displays a BNs involving two subpopulations (Hispanic and Caucasian) and two loci (\texttt{D13} and \texttt{FGA}). The node ``subpopulation identification'' determines the subpopulation from which the founder genes (``suspect paternity'', ``suspect maternity'', ``offender paternity'' and ``offender maternity'') are drawn. The suspect/offender and corresponding paternity/maternity gene form the generic founder module as mentioned earlier in this section. Whether the trace genotype at \texttt{D13} and \texttt{FGA} takes values from the suspect or the offender genotypes depends on the node ``suspect is guilty''. 

\subsubsection{DNA mixtures}
\label{sec: DNA mixture graphical models}
\begin{figure*}[t]
    \centering
    \subfigure[OOBN for a two-contributor DNA mixture.]{
        \begin{minipage}{9cm}
            \centering
            \includegraphics[width=1\textwidth]{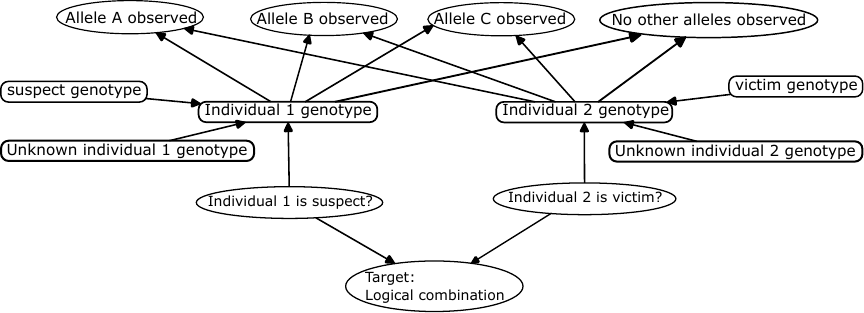}
        \end{minipage}
    }~
    \subfigure[DNA mixture with peak height information \citep{cowell2004identification}.]{
        \begin{minipage}{9cm}
            \centering
            \includegraphics[width=1\textwidth]{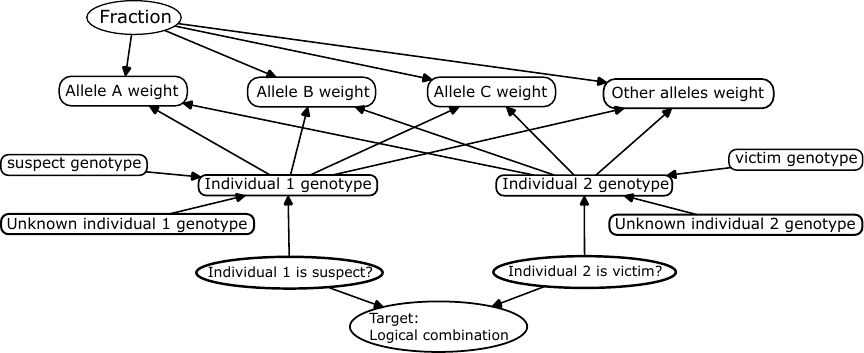}
        \end{minipage}
    }
    \caption{\small\textbf{OOBNs for DNA mixtures.} \textbf{(a)} OOBN with three nodes representing the detected alleles A, B, and C, and the node ``no other alleles observed'' representing the hypothesis of A, B, C being the only detected alleles. The true or false of these four nodes depends on the genotypes of both contributors, individuals 1 and 2, whose nodes are generic identification modules, as they take values from the generic founder modules \textit{suspect/unknown individual 1 genotype} and \textit{victim/unknown individual 2 genotype}, respectively, based on whether \textit{individual 1 is the suspect} and \textit{individual 2 is the victim}. The target node represents the four logical combinations of these last two nodes. \textbf{(b)} OOBN with almost the same structure of the one in (a), except that the evidence nodes contain information about peak height of the electropherogram, and the additional node \textit{fraction} represents the proportion of DNA contributed by the two individuals across all markers.}
    \label{fig: OOBN DNA mixture}
\end{figure*}

In Section~\ref{sec: DNA mixture likelihood framework}, we discussed the analysis of DNA mixtures via LR. Here, we present methods based on BNs. In the case where two individuals contributed to a DNA sample, typically the two competing hypotheses are:
\begin{itemize}
\item[$H_{\rm p}$:] The victim and the suspect contributed to the mixture.\vspace{-3mm}
\item[$H_{\rm d}$:] The victim and an unknown individual (unknown individual 1) contributed to the mixture. 
\end{itemize}
If the victim is not the main source of the DNA sample, we should also consider one more unknown individual, and consider the additional hypotheses
\begin{itemize}
\item[$H_{\rm p'}$:] The suspect and an unknown individual (unknown individual 2) contributed to the mixture. \vspace{-3mm}
\item[$H_{\rm d'}$:] Two unknown contributors (unknown individual 1 and 2) contributed to the mixture.
\end{itemize}
Suppose that exactly three alleles A, B, and C, are identified in a mixed DNA sample, and that the genotypes of the victim and the suspect are observed. In the OOBN in Figure~\ref{fig: OOBN DNA mixture}(a), three nodes represent the detected alleles A, B, and C, and the node ``no other alleles observed'' represents the hypothesis of A, B, C being the only detected alleles. The true or false of these four nodes depends on the genotypes of both contributors, individuals 1 and 2, whose nodes are generic identification modules, as they take values from the generic founder modules \textit{suspect/unknown individual 1 genotype} and \textit{victim/unknown individual 2 genotype}, respectively, based on whether \textit{individual 1 is the suspect} and \textit{individual 2 is the victim}. The target node represents the four logical combinations of these last two nodes, corresponding to the hypotheses $H_{\rm p},H_{\rm p'},H_{\rm d},H_{\rm d'}$. 

As mentioned in Section~\ref{sec: DNA mixture likelihood framework}, peak heights or peak areas in the electropherogram (Section~\ref{sec: DNA}) can also be exploited in mixed DNA typing. Since the proportion of DNA contributed by each individual is constant across all markers, we can use a common \textit{fraction} module representing the proportion of DNA contributed by the two individuals across all markers, as shown in Figure~\ref{fig: OOBN DNA mixture}(b), which is an extension of Figure~\ref{fig: OOBN DNA mixture}(a) where evidence nodes contain information about peak height of the electropherogram \citep{cowell2004identification,dawid2018graphical}.

The presented OOBNs can also be augmented with additional modules to account for potential mutation and issues such as allele drop-out, and to model the uncertainty on the number of contributors in mixed DNA samples \citep{mortera2003probabilistic}.

\subsection{Chain Event Graphs}
\label{sec: Chain event graph (CEG)}
Chain Event Graphs (CEG) were first introduced by \cite{smith2008conditional}. CEGs can be used to analyze causal hypotheses in domains with discrete, asymmetric, and context-specific information, where the measurement variables can exhibit varying sets of potential outcomes based on different combinations of values for sets of ancestral variables which cannot be accommodated by regular BNs \citep{churchill1995accurate,bedford2001probabilistic,french1989readings,lyons1990random,boutilier2013context,madrigal2004causal,poole2003exploiting}, as well as graphs with structural missing values \citep{walley2023cegpy}. In brief, a CEG is an \textit{event tree} \citep{shafer1996art}, which is a directed, rooted tree, and is very commonly used for activity-level hypotheses (Section~\ref{sec: Likelihood ratio framework}). The paths from the \textit{root} to the \textit{leaves} (or terminal nodes) represent chains of events and provide labels for the various potential developments of the described process. The non-leaf vertices are called \textit{situations} and are the indices of random variables that characterize the subsequent stages of potential developments in the unfolding process. Let us consider an example, which is an adaptation of a fictional crime by \cite{thwaites2010causal}.\vspace{-3mm}

\paragraph{Example: ``Vandalism''}
The police arrests and detains a suspect believed to have thrown a brick through the window of a shop. Multiple glass fragments are collected from the suspect's jacket and from the broken window. While the police aim to take the suspect to court, there might be reasons hindering the proceedings.

\begin{figure*}[t]
    \centering
    \includegraphics[width=0.6\textwidth]{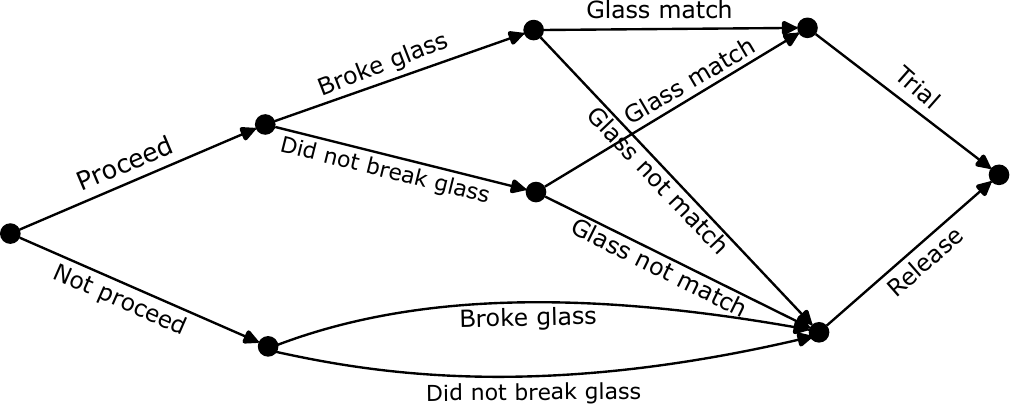}
    \caption{\small\textbf{CEG for the ``Vandalism'' example.} The root on the left side is the node where a decision on whether to proceed with the case is taken. Whether the suspect broke the glass does not depend on whether the case proceeds or not; thus arrows ``broke glass'' and ``did not break glass'' are present in both situations ``proceed'' and ``not proceed.'' If the police decides to proceed, the glass evidence is sent to the forensic laboratory no matter whether the suspect is actually guilty or not. If the glass evidence matches, the suspect is brought to trial. Otherwise, if the glass evidence does not match, the suspect is released. Same outcome is obtained if the police decides not to proceed at all with the case.}
    \label{fig: CEG example}
\end{figure*}

In Figure~\ref{fig: CEG example}, we show a CEG describing this case. The root on the left side is the node where a decision on whether to proceed with the case is made. Whether the suspect broke the glass does not depend on whether the case proceeds or not; thus arrows ``broke glass'' and ``did not break glass'' are present in both situations ``proceed'' and ``not proceed.'' If the police decides to proceed, the glass evidence is sent to the forensic laboratory no matter whether the suspect is actually guilty or not. If the glass evidence matches, the suspect is brought to trial. Otherwise, if the glass evidence does not match, the suspect is released. Same outcome is obtained if the police decides not to proceed at all with the case. A total of six chains of events (directed paths from the root to the leaf) can be identified in the graph. By assigning conditional probabilities to the outcomes (arrows) of every situation (non-leaf vertex), the probability of each of the six chains of events can be computed as the product of the related conditional probabilities.

\section{Conclusion}
\label{sec: Conclusion}
Statistics plays a crucial role in the analysis, interpretation, and uncertainty quantification of forensic evidence. However, national-level standards that could ensure the appropriate application of statistics in forensic investigations have not yet been established. The obstacles in promoting statistics in court are multifaceted, and include: difficulties arising in the application of complex statistical models; challenges due to subjectivity and cognitive bias often affecting evidence examination; and the limited understanding of statistical concepts by laypeople involved in the decision making process within legal proceedings. These issues have had serious consequences in the functionality of the U.S.~criminal justice system, as emphasized by the numerous cases of wrongful conviction of innocent individuals.

Likelihood ratios are commonly computed to quantify the support of a prosecutor's hypothesis over the defense's hypothesis, based on the available forensic evidence. The application of likelihood ratios in forensic science has evolved over time, in terms of sophistication of probabilistic models and verbal interpretation, together with advancements in evidence collection technology, notably DNA typing. However, challenges arise in constructing likelihood ratios for various types of evidence, such as pattern evidence like fingerprints, because of the limited prior knowledge regarding the population distribution of evidence.

Graphical models allow us to represent multiple hypotheses and evidence as nodes of a graph connected by edges or arrows that signify statistical dependence and causality. These models let us compute conditional probabilities and likelihood ratios from intricate scenarios involving multiple types of evidence simultaneously. Special families of graphs, including object-oriented Bayesian networks and chain event graphs, are particularly useful to scrutinize the components of complex forensic investigations. Object-oriented Bayesian networks have flexible hierarchical structures that can easily adapt to disparate forensic investigations. These models are excellent for DNA typing in both criminal and civil paternity cases, based on single or multiple contributor samples, and can be designed in accordance with a variety of assumptions on founder genes. Chain event graphs can be used to analyze causal hypotheses in domains with discrete, asymmetric, and context-specific information, where the measurement variables can exhibit varying sets of potential outcomes based on different combinations of values for sets of ancestral variables.

Beyond the approaches discussed in this article, statistical science offers several other methods for the analysis of forensic evidence, including machine learning techniques \citep{richmond2020ai, jadhav2020artificial, kowalski2023forensic, gupta2020artificial} for the analysis of polygraph data \citep{constancio2023deception}, digital forensics \citep{tallon2014data, iqbal2020advancing, oladipo2020state,qadir2021applications,ngejane2021digital}, pattern evidence \citep{carriquiry2019machine}, forensic autopsy \citep{yeow2014application}, and DNA typing \citep{katsara2021evaluation, liu2020forensic}.

The rigorous application of statistics in forensic investigations is pivotal for ensuring justice. The continuous advancement of statistical techniques for evidence assessment holds the promise of delivering more just outcomes within the criminal justice system. However, this can be made possible only if appropriate standards for the accurate and equitable application of statistics in the courtroom are established at the national-level. While certain standards, such as the use of 20 STR loci for DNA identification \citep{karantzali2019effect}, have been established and implemented, there remains a lack of rules for the analysis of other types of evidence, particularly pattern evidence like fingerprints and marks. National-level guidelines must be established to avoid situations where the implementation of statistical methods and the interpretation of results are left to subjectivity and arbitrariness, potentially yielding wrongful convictions. These guidelines should include: (a) criteria for determining thresholds for likelihood ratios and levels of statistical significance depending on the seriousness of a crime; (b) rules to limit examiners' access to contextual information beyond the evidence sample object of analysis, thereby reducing the risk of cognitive bias; (c) rules that identify the situations where the involvement of professional statisticians as expert witnesses is required, independently of prosecutor's or defendant's requests. Without clear guidelines, issues can easily arise at any stage of a legal process, from evidence collection to sentencing.

\bibliographystyle{apalike}
\bibliography{sample}

\end{document}